\documentclass[letterpaper, 10 pt, conference]{ieeeconf}  
\IEEEoverridecommandlockouts                             
\usepackage{comment}
\usepackage{mathbbol}
\usepackage{tabularx}
\usepackage{xcolor, soul}
\sethlcolor{green}

\let\labelindent\relax

\usepackage{hyperref}    % Hyperlinks
\hypersetup{
    colorlinks=true,      % Enable colored links
    linkcolor=blue,       % Internal links color
    citecolor=blue,       % Citation links color
    filecolor=magenta,    % File links color
    urlcolor=cyan         % External URLs color
}

\usepackage{etoolbox}
\makeatletter
\patchcmd{\@makecaption}
  {\scshape}
  {}
  {}
  {}
\makeatother
\usepackage[caption=false]{subfig}

\usepackage{euscript}[mathcal]
\usepackage[inline]{enumitem}
\usepackage{multirow}
\usepackage{amsmath}
\usepackage{esvect}
\usepackage{color}
\usepackage{subfig}
\usepackage{url}

\definecolor{newcolor}{rgb}{0.858, 0.188, 0.478}
\usepackage[english]{babel}
\usepackage[caption=false]{subfig}
\usepackage{multirow}
\usepackage{wrapfig}
\usepackage{booktabs}
\usepackage{array}

\usepackage{stfloats}
\usepackage[utf8]{inputenc}
\usepackage[T1]{fontenc}
\usepackage{amsmath}

\usepackage{graphicx}
\usepackage{adjustbox}
\graphicspath{{./images/}}
\usepackage{amssymb}
\usepackage{longtable}
\usepackage{float}
\usepackage{algorithm}
\usepackage{algpseudocode}
\usepackage[font=normalsize, labelfont=bf, justification=centering]{caption}
\usepackage{booktabs} 
\usepackage{makecell}
\makeatletter 
\@namedef{ver@float.sty}{3000/12/31}
\makeatother

\usepackage{mathptmx} % assumes new font selection scheme installed

\title{\LARGE \bf
Mode Choice Heterogeneity Among Zero-Vehicle Households: A Latent Class Cluster Approach}

\author{
Nancy Kasamala\textsuperscript{*}, Arthur Mukwaya\textsuperscript{*}, Nana Kankam Gyimah\textsuperscript{*},
Judith Mwakalonge\textsuperscript{*}, \\ 
Gurcan Comert\textsuperscript{+}, Saidi Siuhi\textsuperscript{*}, Akinbobola Jegede\textsuperscript{*} \\
{\small \textsuperscript{*} South Carolina State University, Orangeburg, South Carolina, US, 29117} \\
{\small \textsuperscript{+} North Carolina A\&T State University, Greensboro, North Carolina, US, 27411}
}

\begin{document}
\maketitle
%%%%%%%%%%%%%%%%%%%%%%%%%%%%%%%%%%%%%%%%%%%%%%%%%%%%%%%%%%%%%%%%%%%%%%%%%%%%%%%%
\begin{abstract}
In transportation planning, Zero-Vehicle Households (ZVHs) are often treated as a uniform group with limited mobility options and assumed to rely heavily on walking or public transit. However, such assumptions overlook the diverse travel strategies ZVHs employ in response to varying trip needs and sociodemographic factors. This study addresses this gap by applying a weighted Latent Class Cluster Analysis (LCCA) to data from the 2022 National Household Travel Survey (NHTS) to uncover distinct mobility patterns within the ZVH population. Using travel mode and trip purpose as indicators and demographic, economic, and built environment variables as covariates, we identified three latent classes:Shared mobility errand workers (36.3\%), who primarily use transit and ridehailing for commuting and essential activities; car based shoppers (29.9\%), who depend on informal vehicle access for longer discretionary trips and active travel Shoppers (33.8\%), who rely on walking or cycling for short, local shopping oriented travel. These behavioral findings enable policymakers to develop differentiated planning solutions to the specific needs of each segment among the ZVHs population across varied geographic and demographic settings.\\

\end{abstract}
\vspace{1em}

\noindent\textbf{Keywords:} Zero-Vehicle Households, Latent Class Cluster Analysis, Heterogeneity, Mode Choice, Transportation Planning.

%%%%%%%%%%%%%%%%%%%%%%%%%%%%%%%%%%%%%%%%%%%%%%%%%%%%%%%%%%%%%%%%%%%%%%%%%%%%%%%%

\section{Introduction}

Understanding mode selection patterns is an important aspect of transportation planning and policy formulation, particularly in efforts to promote mobility and improve accessibility. A unique and policy-relevant subpopulation is zero-vehicle households (ZVHs), those that do not own or lease a vehicle. These households comprise of 8.4\% of all U.S. households, according to the 2022 National Household Travel Survey (NHTS)~\cite{nhts2023households}. ZVHs often rely on a range of non-driving alternatives, including public transit, walking, cycling, and shared mobility \cite{Suzukawa2024, Ghimire2024}. However, their travel patterns are shaped by a different mix of socioeconomic factors, lifestyle choices, access limitations, and regional contexts \cite{van2023zero}. As urbanization increases, understanding how ZVHs make travel decisions is critical for guiding inclusive and effective transportation planning strategies.

Despite the availability of nationally representative travel data, most of the existing research treats ZVHs as a homogeneous group \cite{tomer2011transit}. This is a critical limitation as these households exhibit significant behavioral variation in terms of trip purpose, mode selection, and access to transport options. There is a growing need for data-driven approaches that can uncover hidden population subgroups and provide deeper behavioral insights \cite{oecd2023household}. In particular, segmentation techniques that do not rely on predefined outcomes or assumptions about utility structures are needed to identify meaningful patterns and support more responsive transport planning.

This study addresses the core problem that current travel behavior models lack the ability to identify and explain latent behavioral heterogeneity within ZVHs nationally. Most conventional approaches aggregate over diverse traveler profiles, limiting their ability to inform targeted interventions. Addressing this issue is methodologically challenging due to the following reasons:

\begin{enumerate}
    \item Prior studies often have limited geographic scope, which constrains the generalizability of their findings across different community types and travel environments.
    
    \item While the NHTS provides a rich dataset, its structure does not inherently reveal behavioral segments, requiring the application of specialized unsupervised methods to extract deeper insights.
    
    \item Mode choice behavior within ZVHs is shaped by a complex and multidimensional set of factors including sociodemographic characteristics, trip features, and contextual variables which are not adequately captured by aggregate or single-equation models.
\end{enumerate}

While previous studies have explored aspects of carless travel behavior, many rely on traditional regression \cite{wang2025rural} techniques or focus on descriptive statistics \cite{klein2024invisible}. These methods, although informative, fall short in revealing hidden behavioral structures that emerge from large-scale survey data. Although there has been increasing research at the regional level recognizing the heterogeneity within ZVHs, such studies are constrained by geographic specificity~\cite{patwary2025travel,brown2017car}, limiting the generalizability of their findings to the national level. Meanwhile, national-level studies have begun to examine travel behavior among zero-vehicle populations~\cite{Li03072025}, but many of these efforts remain limited in demographic scope or rely on fixed subgroup categorizations rather than data-driven segmentation. Consequently, there remains a methodological and empirical gap in capturing the full behavioral diversity of ZVHs using scalable, interpretable clustering models.

To overcome these limitations, this study applies a \textit{Latent Class Cluster Analysis (LCCA)} framework to the 2022 NHTS dataset. LCCA is an unsupervised, model-based clustering approach that enables the identification of latent subgroups based on observed categorical travel behavior indicators, such as mode choice and trip purpose. The approach does not require explicit assumptions about rational utility or decision-making structures, making it particularly well-suited for uncovering underlying heterogeneity. Additionally, by incorporating sociodemographic and contextual covariates, the analysis allows for the interpretation of class membership characteristics, supporting policy-relevant insights.

The key contributions of this study are as follows:

\begin{enumerate}
    \item This study develops a nationwide LCCA framework to reveal distinct behavioral segments among zero-vehicle households, overcoming geographic and methodological limitations in prior research.

    \item Using categorical travel indicators, the model uncovers latent travel behavior patterns and links them with socioeconomic and geographic covariates to provide a detailed profile of each behavioral class.

    \item The study translates complex survey data into interpretable, actionable traveler profiles that can inform targeted transportation policies at the national scale.
\end{enumerate}

The remainder of this paper is organized as follows: Section~\ref{sec:Related Works} reviews prior studies on travel mode choice modeling, ZVHs, and latent class-based segmentation. Section~\ref{sec:Proposed Methodology} outlines the proposed methodology, including data preparation, feature engineering, and descriptive analysis of travel behavior among ZVHs. Section~\ref{sec:lcca_framework} details the Latent Class Cluster Analysis (LCCA) framework, including model structure, estimation procedures, and indicator-covariate integration. Section~\ref{Section:Results and Discussion} presents the empirical findings, including latent class segmentation and profiling. Section~\ref{sec:discussion} interprets the results implications, limitation and future works for transportation planning and policy. Finally, Section~\ref{sec:conclusion} provides a summary of our research work.

\section{Related Works}
\label{sec:Related Works}
ZVHs represent a critical segment of travelers influenced by unique mobility constraints such as economic limitations, infrastructure availability, and varying personal preferences. Traditional modeling techniques, notably multinomial logit and regression analyses, tend to treat this group as homogeneous, overlooking substantial internal diversity. Recently, latent class modeling (LCM) and clustering analyses have emerged as more advanced approaches capable of identifying heterogeneous traveler segments based on underlying preferences and behaviors. Literature reviewed herein falls broadly into two categories: explicit studies on ZVHs, and latent-class based clustering analyses.

\subsection{Empirical Research on Zero-Vehicle Households}
A number of empirical studies have directly explored the behavior, needs, and coping strategies of ZVHs. Klein et al.~\cite{klein2024invisible} conducted a Descriptive statistics, thematic analysis, and comparative validation study in Baltimore, Maryland, to examine how car-less households access transportation for daily needs. They found that ride-hailing was commonly used for grocery and leisure trips, though the study did not examine modal frequency or the role of micro-mobility. In the study done by Brown~\cite{brown2017car}, using the 2012 California Household Travel Survey, differentiated between ``car-less'' and ``car-free'' households, finding that voluntarily car-free individuals traveled more and relied more on transit and active modes. By utilizing census-like microdata from Flanders, Belgium, Van Eenoo~\cite{van2023zero} profiled the socio-demographic composition of ZVHs, identifying vulnerable groups such as low-income urban singles and highlighting rural mobility barriers. 

Complementing these quantitative findings, qualitative studies offer deeper contextual insight. Lagrell et al.~\cite{lagrell2018accessibility} used in-depth interviews to understand how car-less families in Gothenburg managed daily travel, revealing a reliance on proximity, cycling, and walking, especially for mandatory trips. Similarly, Paijmans and Pojani~\cite{paijmans2021living} investigated voluntary car-free living in Brisbane, Australia, identifying motivations such as environmental concern and health, and noting the importance of public transit, social support, and flexible work arrangements in sustaining car-free lifestyles. Using 2012 California Household Travel Survey data, Mitra and Saphores~\cite{mitra2018carless} compared mode use and trip distances between carless and motorized households, revealing notable differences but offering limited insight into behavioral diversity. By applying descriptive statistics, logistic regression, and GIS mapping to national datasets including the American Community Survey (ACS) and National Household Travel Survey (NHTS), Wang et al.~\cite{wang2025rural} examined the rural–nonrural divide in the U.S., highlighting significant unmet travel needs among rural car-less populations, though they did not segment households further to explore behavioral heterogeneity. In Sweden, Lagrell and Gil Solà~\cite{lagrell2021car} analyzed national travel survey data to explore when and why nominally car-less individuals still use cars. Their findings showed that residual car use is common in social trips and often occurs through shared or borrowed vehicles, underscoring the role of situational constraints and the need for shared mobility solutions.

\subsection{Latent-Class Based Clustering Analyses}
Recently, latent class cluster analysis (LCCA) and related approaches have emerged as powerful methodologies to identify traveler heterogeneity through segmenting behaviors based on latent preferences. Iogansen et al.~\cite{iogansen2025ridehailing} conducted a latent-class cluster analysis using GPS-based one-week travel diaries from multiple Californian metros. The study notably classified travelers based on their trip purposes (work and non-work-related) and mode choice behaviors (single-occupant vehicles, carpool, transit, biking, and walking), identifying four distinct traveler segments (drive-alone, carpoolers, transit users, and cyclists). The research emphasized that transit users exhibited the highest multimodality and frequent use of ride-hailing services. 
LCCA framework was also used by Wang et al.~\cite{wang2022identifying} to identify distinct mobility preference segments among low-income individuals in Detroit and Ypsilanti, Michigan, categorizing them as bus-oriented, ridehailing-oriented, or mobility-on-demand transit-oriented, and revealed critical interactions between income, vehicle access, and trip purposes. Similarly, Circella et al.~\cite{circella2023mode} utilized latent class cluster analyses to identify mode substitution patterns triggered by ride-hailing adoption across several U.S. cities, showcasing different behavioral responses among latent traveler segments, though without incorporating actual GPS-based validation of travel patterns. In the Netherlands, Meester et al.~\cite{meester2024framework} used Latent Class Logit modeling, Monte Carlo simulation, and GIS-based accessibility analysis to examine how socio-demographic factors affect job accessibility in low-car and car-free development zone. The findings revealed that accessibility outcomes varied notably across latent classes, with lower-income and less mobile groups experiencing reduced job access despite similar geographic proximity. Other notable studies have also employed latent class and clustering approaches to examine travel behavior heterogeneity across diverse populations and settings, further reinforcing the methodological relevance of data-driven segmentation in transportation research \cite{Ton2020, Rasmussen2023, Molin2016,Lee2020}

Despite significant advancements, several gaps and limitations persist across reviewed literature. First, there remains an over-reliance on descriptive methods with limited predictive power for informing targeted policy interventions. Secondly, geographic limitations restrict the external validity and generalizability of findings. Finally, many studies reviewed utilized pre-pandemic datasets, missing recent trends like the surge in micromobility options (e-scooters, shared bicycles) and evolving travel behaviors post-COVID-19. 

To overcome these limitations, the current research applies Latent Class Choice Analysis (LCCA) to a nationally representative dataset ,NHTS capturing diverse geographic contexts, sociodemographic variables, and contemporary modal preferences. This approach allows identification of behavioral segments within ZVHs, directly addressing existing gaps related to geographic generalizability, methodological rigor, and sociodemographic complexity.

\section{Proposed Methodology}
\label{sec:Proposed Methodology}
\subsection{Data Description and Processing}

This study uses data from the 2022 United States National Household Travel Survey (NHTS) \cite{fhwa2022} that captures detailed information on household, person, and trip characteristics. Household, person, and trip files were merged using house ID, person ID and trip ID that were common across them and filtered to include only households with zero vehicles. Various variables relevant to this study were selected from the merged dataset that included travel characteristics (transport mode, general trip purpose and distance travelled), socio-economic and demographic attributes (age, income, gender, race, employment status and education) and built environment characteristics (population density). All rows containing missing values were removed, and the final sample size consisted of 795 rows.

\subsection{Feature Engineering}
Multi categorical and continuous variables were grouped into new factor variables to ensure meaningful and interpretable class distinctions. Travel mode data were grouped into three broad categories: private Vehicle which includes car, van, SUV/crossover, pickup truck, recreational vehicle, motorcycle, and e-scooters; shared mobility having public or commuter bus, school bus, streetcar or trolley, subway or elevated rail, commuter rail, Amtrak, taxicab or limo service, other ride-sharing services, and paratransit/dial-a-ride; and active travel which comprises bicycle and walking modes. Age was grouped into five life stage categories: 0--17, 18--30, 31--44, 45--65, and over 64. Continuous trip distances (miles) were binned into four ranges: 0--1, >1--3, >3--10, and >10. Race was simplified into four categories: white, black or african american, asian, and others. Education levels were grouped into four groups: up to high school, some college or trade school, bachelor's degree, and advanced degree. Population density (persons per square mile) was categorized into three levels: 0--999, 1{,}000--9{,}999, and over 10{,}000. Household income was grouped into three brackets: \$0--24{,}999, \$25{,}000--74{,}999, and above \$74{,}999.

The 7 day national trip weights were normalized to a mean of approximately 10 and a minumum mean of 1 since  flexmix only allows integers. This ensured that the resulting parameter estimates reflect the characteristics of the broader population of ZVHs.

\begin{figure*}[htbp]
\centering
\includegraphics[width=0.8\linewidth]{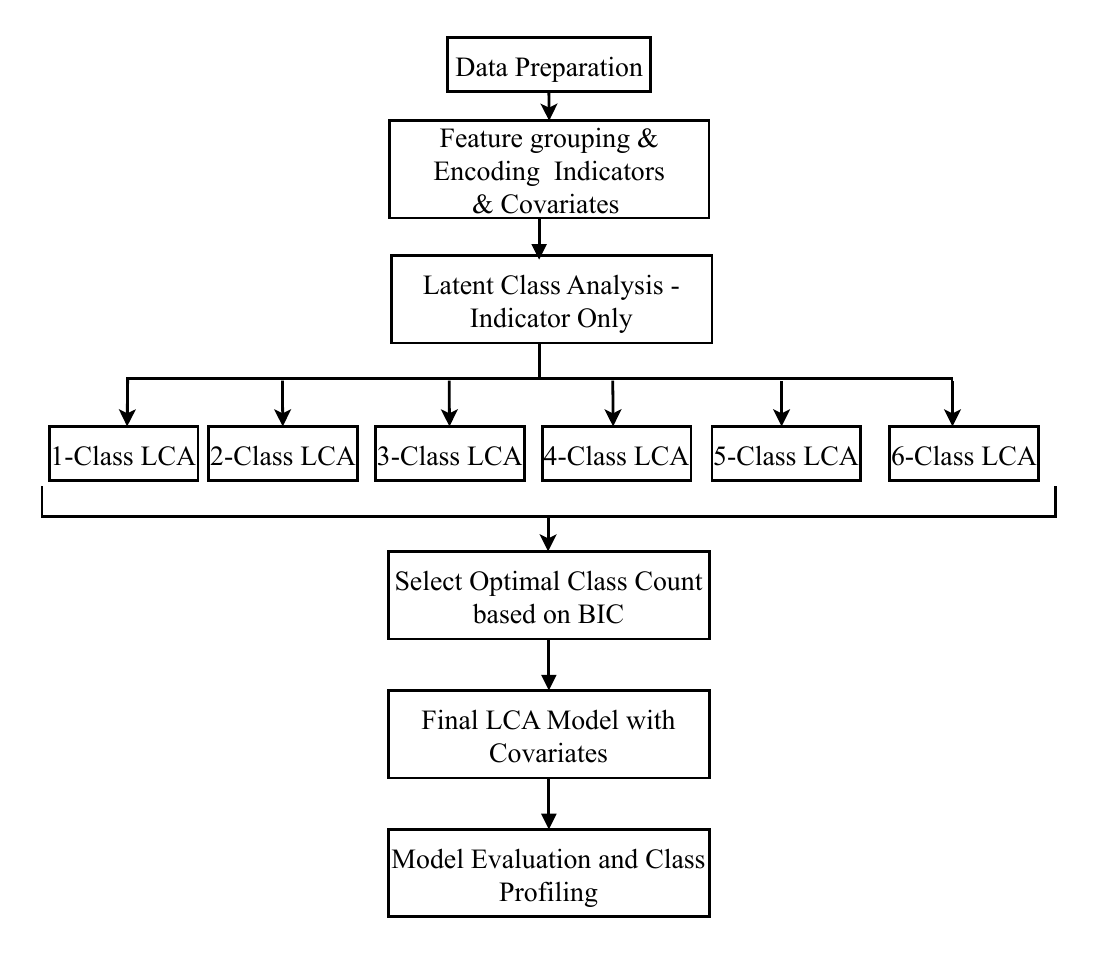}
\caption{Overview of the latent class modeling pipeline, including data preparation, indicator-only LCA estimation, model selection using BIC, covariate-enhanced model fitting, and final class profiling.}
\label{fig:lca_pipeline}
\end{figure*}

Descriptive analysis for the final sample characterizing both the unweighted the weighted estimates was done as shown in Table \ref{tab:descriptive_stats} with percentage distributions for the various variables used in LCCA. Active travel accounts for 48.81\% of unweighted trips but declines to 40.58\% when weighted. private vehicle (29.23\%) and shared mobility (30.19\%) together represent the majority of weighted trips. Home-based shopping is the most frequent trip purpose across both unweighted (28.81\%) and weighted (27.86\%) data while Home-based work and Home-based other increase after weighting. Short trips (0--1 miles) dominate the unweighted data (38.87\%), but longer trips ($>$1--3 mi and $>$10 mi) are more prominent in weighted estimates.Individuals aged 31--44 are the largest group (36.94\% weighted), and females make up a greater share of weighted trips (59.26\%).  White individuals decrease from 69.81\% unweighted to 48.66\% weighted while Black or African American individuals rise to 28.94\%. Most travelers are workers (58.05\%), and educational qualification skews lower with 55.64\% having a high school education or less. Nearly half of all trips are from low-income households (0--\$24,999), and the majority originate in medium or high-density areas for both weighted and unweighted samples. \noindent
To provide a structured overview of the full methodological pipeline, Figure~\ref{fig:lca_pipeline} summarizes the main steps involved in the latent class modeling process, from data preparation to model estimation and evaluation. The subsequent section elaborates on model implementation and evaluation procedures.

\begin{table*}[ht]
    \centering
    \caption{Descriptive statistics of the variables with their weighted and  unweighted percentages}
    \label{tab:descriptive_stats}
    \resizebox{2.0\columnwidth}{!}{%
    \begin{tabular}{>{\raggedright\arraybackslash}p{5cm} >{\raggedright\arraybackslash}p{5cm} rr}
        \toprule
        \textbf{Variable} & \textbf{Category} & \textbf{Unweighted\%} & \textbf{Weighted\%} \\
        \midrule

        \multirow{3}{=}{Mode group} 
            & Private Vehicle & 25.03 & 29.23 \\
            & Shared Mobility & 26.16 & 30.19 \\
            & Active Travel  & 48.81 & 40.58 \\

        \addlinespace

        \multirow{5}{=}{General Trip purpose} 
            & Home-based work& 14.34 & 18.79 \\
            & Home-based shopping& 28.81 & 27.86 \\
            & Home-based social/recreational& 12.70 & 11.62 \\
            & Home-based other& 22.01 & 24.27 \\
            & Not a home-based trip& 22.14 & 17.45 \\

        \addlinespace

        \multirow{5}{=}{Age} 
            & 0--17 & 2.52 & 3.94 \\
            & 18--30 & 22.77 & 23.23 \\
            & 31--44 & 31.82 & 36.94 \\
            & 45--65 & 28.81 & 23.08 \\
            & >64 & 14.09 & 12.82 \\

        \addlinespace

        \multirow{4}{=}{Distance (mi)} 
            & 0--1 & 38.87 & 33.12 \\
            & >1--3 & 27.30 & 29.52 \\
            & >3--10 & 23.90 & 24.07 \\
            & >10 & 9.94 & 13.30 \\

        \addlinespace

        \multirow{3}{=}{Income (\$)} 
            & 0--24,999 & 39.12 & 49.90 \\
            & 25,000--74,999 & 26.29 & 28.51 \\
            & >74,999 & 34.59 & 21.59 \\

        \addlinespace

        \multirow{2}{=}{Gender} 
            & Male & 47.67 & 40.74 \\
            & Female & 52.33 & 59.26 \\

        \addlinespace

        \multirow{4}{=}{Race} 
            & White & 69.81 & 48.66 \\
            & Black or African American & 15.85 & 28.94 \\
            & Asian & 9.94 & 10.74 \\
            & Other race & 4.40 & 11.66 \\

        \addlinespace

        \multirow{2}{=}{Employment status} 
            & Worker & 57.11 & 58.05 \\
            & Not worker & 42.89 & 41.95 \\

        \addlinespace

        \multirow{4}{=}{Education} 
            & Upto Highschool & 26.54 & 55.64 \\
            & Some College or Trade School & 23.27 & 18.13 \\
            & Bachelor's Degree & 32.20 & 17.44 \\
            & Advanced Degree & 17.99 & 8.78 \\

        \addlinespace

        \multirow{3}{=}{Population density (persons per square mile)} 
            & 0--999 & 10.31 & 11.49 \\
            & 1,000--9,999 & 42.89 & 48.64 \\
            & >10,000 & 46.80 & 39.87 \\

        \bottomrule
    \end{tabular}}
\end{table*}

\section{Experimental Setup}
\label{sec:lcca_framework}
LCCA was performed using the flexmix package in the R statistical programming language executed within a Google Colaboratory environment. The package was selected for its flexibility in specifying finite mixture models and the ability to incorporate survey weights, making it well suited for this study in comparison to poLCA.

\subsection{Latent Class Cluster Analysis (LCCA)}

LCCA models presume the existence of unobserved latent classes within a population, where individuals within each class are assumed to be homogeneous with respect to their response patterns on observed indicator variables but heterogeneous across classes. The overall probability of an individual's observed responses is a weighted sum of class- specific probabilities where the weights are the probabilities of belonging to each class which is useful in identifying distinct travel behavior patterns ZCHs.

Let $Y_i = (Y_{i1}, Y_{i2}, \ldots, Y_{iL})$ denote the set of $L$ indicator variables for individual $i$, and $Z_i = (Z_{i1}, Z_{i2}, \ldots, Z_{iJ})$ denote the set of $J$ covariates for individual $i$. Let $C_k$ be the $k$-th latent class, for $k \in \{1, 2, \ldots, K\}$ total classes. The overall probability of observing the response pattern $Y_i$ for individual $i$, given their covariates $Z_i$, is given in Equation \ref{eq:individual_lca}:

\begin{equation}
P(Y_i \mid Z_i) = \sum_{k=1}^{K} P(C_k \mid Z_i) \prod_{l=1}^{L} P(Y_{il} \mid C_k)
\label{eq:individual_lca}
\end{equation}

where:
\begin{itemize}
    \item $Y_i$: Observed travel behaviors.
    \item $L$: Total number of indicator variables.
    \item $Y_{il}$: Observed category of the $l$-th indicator variable.
    \item $Z_i$: Observed demographic and socio-economic characteristics.
    \item $C_k$: The $k$-th unobserved latent class
    \item $K$: Total number of latent classes identified in the model
\end{itemize}

This Equation \ref{eq:individual_lca} comprises two main components: a membership model and a measurement model.

\subsubsection{Measurement Model} This describes the relationships between the observed indicator variables (trip characteristics) and the unobserved latent classes. It specifies the conditional probabilities of observing specific indicator responses given membership in a particular latent class. The indicator variables are independent of each other conditional on latent class membership. For each indicator variable $Y_{il}$ with $M_l$ categories and for an individual $i$ belonging to class $C_k$, the probability of observing category $m$ is modeled using a multinomial logit formulation is shown in Equation \ref{eq:multinomial_logit}:

\begin{equation}
P(Y_{il} = m \mid C_k) = \frac{\exp(\beta_{lm \mid k})}{\sum_{m'=1}^{M_l} \exp(\beta_{lm' \mid k})}
\label{eq:multinomial_logit}
\end{equation}

where:
\begin{itemize}
    \item $\beta_{lm|k}$: Coefficient for category $m$ of indicator $L$ within latent class $k$.
    \item $M_l$: Total number of categories for the $l$-th indicator variable.
    \item $m$: Specific category of that indicator.
\end{itemize}

For identification, the parameter for a chosen reference category is constrained to zero for each indicator within each class.

\subsubsection{Membership Model} This shows how the selected covariates predict an individual's probability of belonging to each latent class. The probability of individual $i$ belonging to class $C_k$ given their covariates $Z_i$ is modelled using a multinomial logit formulation in Equation \ref{eq:class_membership}:

\begin{equation}
P(C_k \mid Z_i) = \frac{\exp\left(\gamma_{k0} + \sum_{j=1}^{J} \gamma_{kj} Z_{ij}\right)}{\sum_{k'=1}^{K} \exp\left(\gamma_{k'0} + \sum_{j=1}^{J} \gamma_{k'j} Z_{ij}\right)}
\label{eq:class_membership}
\end{equation}

where:
\begin{itemize}
    \item $\gamma_{k0}$: Intercept term for latent class $k$.
    \item $Z_{ij}$: Value of covariate $j$ for individual $i$.
    \item $\gamma_{kj}$: Coefficient for covariate $Z_{ij}$ in predicting membership in class $k$.
\end{itemize}

\begin{figure*}[htbp]
    \centering
    \includegraphics[scale=0.85]{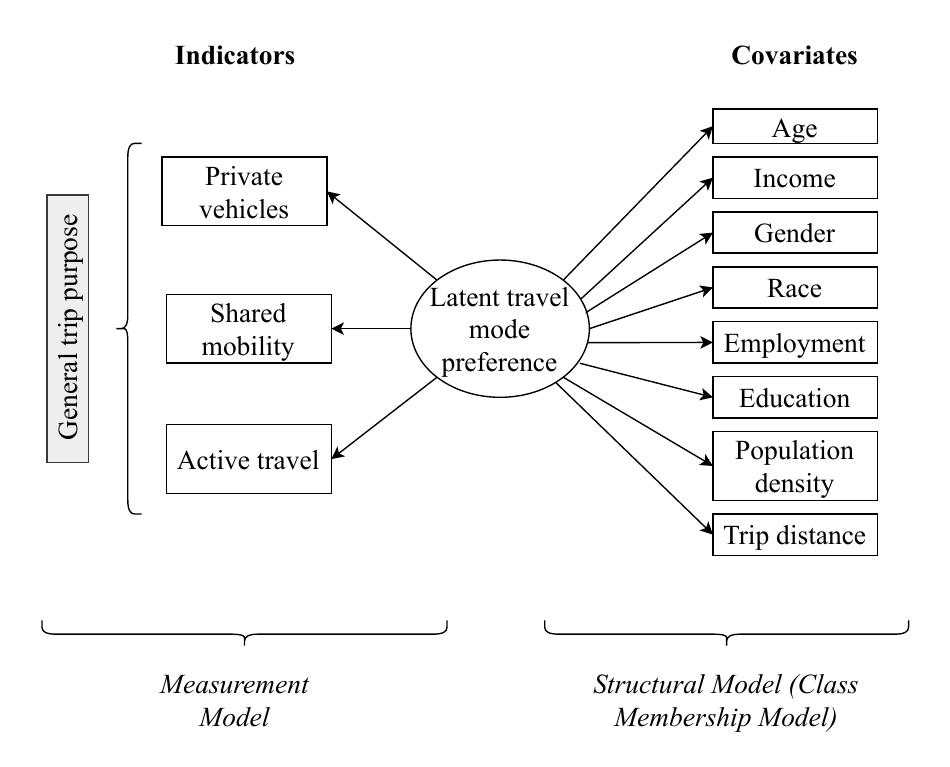}
    \caption{Latent Class Model Structure illustrating the measurement and structural components. Travel mode and general trip purpose serve as indicator variables for classifying latent travel mode preferences, while socio-demographic and contextual covariates predict class membership.}
    \label{fig:lca_structure}
\end{figure*}

Here too, one class is designated as the reference class, and its coefficients are set to zero ($\gamma_{10}=0$ and $\gamma_{1j}=0$ for all $j=1, \ldots, J$).The latent class structure---illustrated in Figure~\ref{fig:lca_structure}---captures how latent mode preferences are revealed through observed indicators and influenced by both individual characteristics and contextual factors.

The LCCA modeling process involved two sequential stages. Initially, LCCA models were fitted using only the following trip characteristics (transport mode, general trip purpose and distance ) as indicators, models with varying numbers of latent classes (K = 1 to 6) were estimated. For each model, multiple random starting points were employed to increase the likelihood of finding the global maximum with a specified maximum number of iterations. The optimal number of classes for LCCA was determined by using goodness-of-fit measures including the Akaike Information
Criterion (AIC), Bayesian Information Criterion (BIC) and log-likelihood.

Upon identifying the optimal number of classes from the indicator-only models, the LCA model was re-estimated to incorporate the covariates in the membership model to explain class assignment. The covariates used are income, age,gender,race, employment status, education and population density. They influence the probability of an individual belonging to a specific latent class.

\subsection{Profiling}
This involved interpreting the identified classes in order to gain insights into the distinct groups with ib ZVHs. The posterior probabilities of each individual belonging to each latent class were obtained. Each observation was then assigned to the latent class for which they exhibited the highest posterior probability of membership. The overall weighted proportions of observations assigned to each class were calculated using survey design methods to provide the population-level prevalence of each identified class of travel behavior. The model's estimated parameters for each indicator within each class were extracted and transformed into probabilities. These probabilities illustrate the expected distribution of each indicator's categories within each latent class which forms the core profile of each class's travel patterns. To understand how the chosen demographic and socio-economic characteristics influenced class membership, a separate multinomial logit model was fitted with normalised weights. In this model, the class assignments from the LCA served as the dependent variable, with the covariates as predictors. Coefficients, Z-values, and p-values were extracted to assess the statistical significance and direction of the relationship between each covariate and the log-odds of belonging to a particular class compared to a reference class.

\section{Results and Discussion}
\label{Section:Results and Discussion}

This section presents the findings from the LCCA detailing the selection of the optimal model, the distinct travel behavior profiles of the identified latent classes based on mode choice and trip purpose, and the influence of socio-economic, demographic, built environment and trip distance covariates on class membership.

\subsection{Latent Class Model Selection}
We estimated latent class models with 1 to 6 classes using goodness-of-fit measures, including AIC, BIC, and log-likelihood. The log-likelihood improved up to the three-class model but showed no further gain beyond that point, suggesting that additional classes did not enhance model fit. While AIC generally improved with additional classes, BIC was prioritized due to its stronger penalty for model complexity and its advantages in model selection for large samples \cite{medel_salgado_bic_aic}. The fit statistics for models with 1 to 6 classes are presented in Table \ref{tab:model_fit}.

\begin{table*}[h]
\centering
\caption{Latent class model fit statistics for classes from 1 to 6}
\label{tab:model_fit}
\resizebox{1.8\columnwidth}{!}{%

\begin{tabular}{lrrrrl}
\toprule
\multirow{2}{*}{\textbf{Classes}} & \multirow{2}{*}{\textbf{AIC}} & \multirow{2}{*}{\textbf{BIC}} & \multirow{2}{*}{\textbf{LogLik}} & \multirow{2}{*}{\textbf{N\_Params}} & \textbf{Class\_Shares} \\
& & & & & \textbf{(per class)} \\
\midrule
1 & 42231.2 & 42273.5 & -21110.6 & 6 & 1 \\
2 & 41364.1 & 41455.5 & -20669.1 & 13 & 0.482, 0.518 \\
3 & \textbf{41255.4} & \textbf{41395.1} & \textbf{-20607.7} & 20 & 0.316, 0.3, 0.384 \\
4 & 41269.4 & 41457.0 & -20607.7 & 27 & 0.259, 0.213, 0.292, 0.236 \\
5 & 41283.4 & 41520.0 & -20607.7 & 34 & 0.165, 0.29, 0.157, 0.181, 0.207 \\
6 & 41297.4 & 41583.1 & -20607.7 & 41 & 0.149, 0.14, 0.156, 0.137, 0.189, 0.227 \\
\bottomrule
\end{tabular}%
}
\end{table*}

The model with three latent classes (K=3) yielded the lowest BIC value (41395.1), indicating the most optimal fit that balances model simplicity and explanatory power. Although the log likelihood continued to slightly improve for models with more classes (K=4 to K=6), the BIC value began to increase, suggesting that the additional complexity did not justify the marginal improvement in fit.

\subsection{Identification of Latent Classes}

The selected three-class model successfully identified distinct subgroups within ZVHs, characterized by differing travel mode preferences and trip purposes. The class sizes were relatively balanced, with Class 1 accounting for 36.3\% of weighted trips, Class 2 for 29.9\% and Class 3 for 33.8\%, indicating meaningful heterogeneity in travel behavior among the ZVH population.

To interpret each class, we examined the model-estimated conditional probabilities of the two indicator variables: mode group and general trip purpose. Table~\ref{tab:conditional_prob} presents these probabilities, which inform the behavioral distinctions between the latent classes.

\begin{table*}[t]
\centering
\caption{Model-Estimated Conditional Probabilities of Indicators by Latent Class}
\label{tab:conditional_prob}
\resizebox{\textwidth}{!}{%
\begin{tabular}{llrrr}
\toprule
\textbf{Indicator} & \textbf{Category} & \textbf{\begin{tabular}[c]{@{}c@{}}Class 1\\ (Shared mobility\\ Errand-workers)\end{tabular}} & \textbf{\begin{tabular}[c]{@{}c@{}}Class 2\\ (Car-based\\ Shoppers)\end{tabular}} & \textbf{\begin{tabular}[c]{@{}c@{}}Class 3\\ (Active travel\\ Shoppers)\end{tabular}} \\
\midrule
\multicolumn{2}{l}{Class Share (\%)} & 36.3\% & 29.9\% & 33.8\% \\
\midrule
\multirow{3}{*}{Mode Group} 
& Private Vehicle        & 0.0894 & \textbf{0.8144} & 0.0491 \\
& Shared Mobility        & \textbf{0.6885} & 0.1389 & 0.0272 \\
& Active Travel          & 0.2221 & 0.0467 & \textbf{0.9237} \\
\midrule
\multirow{5}{*}{Trip Purpose} 
& Home-based work (HBW)                  & \textbf{0.3995} & 0.0821 & 0.0555 \\
& Home-based shopping (HBSHP)           & 0.0679 & \textbf{0.3098} & \textbf{0.4765} \\
& Home-based social/recreational (HBSOC) & 0.0360 & 0.1923 & 0.1351 \\
& Home-based other (HBO)                & 0.3396 & 0.2378 & 0.1396 \\
& Not a home-based trip (NHB)           & 0.1570 & 0.1780 & 0.1932 \\
\bottomrule
\end{tabular}%
}
\end{table*}

\subsubsection{Class 1. Shared Mobility Errand-Workers}  
This class exhibits a pronounced preference for shared transportation services (68.85\%) complemented by moderate active travel usage (22.21\%), while showing minimal private vehicle utilization (8.94\%). The travel pattern centers around home-based work (39.95\%) and home-based other (33.96\%), indicating reliance on public transit systems or ride-hailing platforms for regular commuting and mandatory errand completion.

\subsubsection{Class 2. Car-Based Shoppers}  
Private vehicle usage characterizes this segment (81.44\%) with substantially reduced dependence on shared mobility (13.89\%) and active transportation (4.67\%). Trip purposes are diverse, including home-based shopping (30.98\%), home-based social/recreational (19.23\%), and home-based other (23.78\%), suggesting broad travel needs enabled by access to private vehicles through borrowing, renting or carpooling. The relatively low rate of commuting trips (8.21\%) further supports the discretionary nature of this group.

\subsubsection{Class 3. Active Travel Shoppers}  
Walking and cycling dominate this group's mobility choices (92.37\%), with negligible use  motorized alternatives. The dominant trip purposes are home shopping (47. 65\%) and not a home trip (19. 32\%). This pattern suggests neighborhood-oriented individuals who prefer walking or cycling likely due to proximity to destinations or lack of alternatives.

\subsection{Segment Profiles}
The latent class membership model examines how various socio-demographic, built environment, and travel distance covariates influence the likelihood of an individual belonging to Class 2 or Class 3, relative to Class 1 (the reference class). The estimated coefficients and Z-values are presented in Table \ref{tab:class_membership}. 

\begin{table*}[t]
\centering
\caption{Coefficients from the Class Membership Model (Reference Class: Shared Mobility Errand-Workers)}
\label{tab:class_membership}
{\small
\begingroup
\setlength{\tabcolsep}{5pt}  % Slightly reduced if needed
\renewcommand{\arraystretch}{1.2}
\begin{tabular}{p{3.5cm}p{3.5cm}rrr rrr}
\toprule
\multirow{2}{*}{\textbf{Variable}} & \multirow{2}{*}{\textbf{Category}} 
& \multicolumn{3}{c}{\textbf{Car-Based Shoppers}} 
& \multicolumn{3}{c}{\textbf{Active Travel Shoppers}} \\
\cmidrule(lr){3-5} \cmidrule(lr){6-8}
 & & Coefficient & z-value & Signif. & Coefficient & z-value & Signif. \\
\midrule
\textbf{(Intercept)} & -- & -4.0558 & -5.3145 & *** & 7.5019 & 8.5676 & *** \\
\midrule
\multirow{4}{*}{\textbf{Distance (mi)}} 
& 0--1 & -- & -- & -- & -- & -- & -- \\
& >1--3 miles & 5.4735 & 9.3179 & *** & -13.3994 & -23.4979 & *** \\
& >3--10 miles & 1.6365 & 2.9447 & *** & -21.1467 & -22.3693 & *** \\
& >10 miles & 14.9990 & 17.5324 & *** & -18.3531 & -18.9901 & *** \\
\midrule
\multirow{3}{*}{\textbf{Income (\$)}} 
& 0--24,999 & -- & -- & -- & -- & -- & -- \\
& 25k--74,999 & -0.4120 & -1.9498 & ** & -6.5924 & -14.7082 & *** \\
& >74,999 & 8.9629 & 15.3235 & *** & -1.8213 & -3.8547 & *** \\
\midrule
\multirow{5}{*}{\textbf{Age}} 
& 0--17 & -- & -- & -- & -- & -- & -- \\
& 18--30 & 1.6397 & 3.1939 & *** & 0.9557 & 2.6800 & *** \\
& 31--44 & 2.5474 & 4.4722 & *** & 5.8665 & 13.0189 & *** \\
& 45--65 & 6.4381 & 18.1219 & *** & -0.4980 & -1.0988 & -- \\
& >64 & 5.1646 & 9.6338 & *** & 0.2937 & 2.7275 & *** \\
\midrule
\multirow{2}{*}{\textbf{Gender}} 
& Male & -- & -- & -- & -- & -- & -- \\
& Female & 6.7070 & 14.8009 & *** & -4.4235 & -5.7624 & *** \\
\midrule
\multirow{4}{*}{\textbf{Race}} 
& White & -- & -- & -- & -- & -- & -- \\
& Black or African American & -12.0825 & -22.3750 & *** & -3.0426 & -10.8664 & *** \\
& Asian & -4.8208 & -8.9274 & *** & -4.3697 & -8.0902 & *** \\
& Other race & -3.0603 & -6.6000 & *** & -3.3203 & -7.1610 & *** \\
\midrule
\multirow{2}{*}{\textbf{Employment Status}} 
& Worker & -- & -- & -- & -- & -- & -- \\
& Not Worker & 10.1539 & 22.4044 & *** & 3.8183 & 4.9736 & *** \\
\midrule
\multirow{4}{*}{\textbf{Education}} 
& Upto Highschool & -- & -- & -- & -- & -- & -- \\
& Some College / Trade School & -0.2295 & -0.8197 & -- & 4.0178 & 8.8654 & *** \\
& Bachelor’s Degree & -6.6049 & -12.2318 & *** & 7.3039 & 9.5126 & *** \\
& Advanced Degree & -0.6139 & -1.3242 & -- & 10.2637 & 10.6194 & *** \\
\midrule
\multirow{3}{*}{\makecell[l]{\textbf{Population Density} \\ \textbf{(persons/sq. mile)}}}
& 0--999 & -- & -- & -- & -- & -- & -- \\
& 1,000--9,999 & -7.9292 & -16.3181 & *** & 1.8142 & 3.1818 & *** \\
& >10,000 & -17.0000 & -22.1429 & *** & -0.0906 & -0.0958 & -- \\
\bottomrule
\multicolumn{8}{p{0.95\textwidth}}{\textit{Note:} Blank cells indicate reference categories used in the model for which no coefficients are estimated. “--” denotes these reference categories. *, **, and *** represent statistical significance at the 10\%, 5\%, and 1\% levels, respectively} \\
\end{tabular}
\endgroup
}
\end{table*}

The weighted class profiles in Table \ref{tab:weighted_class_profiles} reveal different patterns of transportation behavior and their associated demographic characteristics.

\begin{table*}[htbp]
\centering
\caption{Summary of Statistics of Zero-Vehicle Household Travel Behavior}
\label{tab:weighted_class_profiles}

\resizebox{2.0\columnwidth}{!}{%
\begin{tabular}{@{}llcccc@{}}
\toprule
\textbf{Variable} & \textbf{Category} & \textbf{Class 1} & \textbf{Class 2} & \textbf{Class 3} & \textbf{Sample} \\
& & \textbf{Shared Mobility} & \textbf{Car-Based} & \textbf{Active Travel} & \textbf{(\%)} \\
& & \textbf{Errand-Workers} & \textbf{Shoppers} & \textbf{Shoppers} & \\
\midrule
\textbf{Class Share} & & \textbf{36.30} & \textbf{29.90} & \textbf{33.80} & \textbf{100.00} \\
\midrule
\multirow{3}{*}{\textbf{Transportation mode}} 
& Private Vehicle & 8.94 & \textbf{81.44} & 4.91 & 31.04 \\
& Shared Mobility & \textbf{68.85} & 13.89 & 2.72 & 30.39 \\
& Active Travel & 22.21 & 4.67 & \textbf{92.37} & 38.57 \\
\midrule
\multirow{5}{*}{\textbf{Age}} 
& 0-17 & 5.13 & \textbf{5.72} & 1.12 & 4.18 \\
& 18-30 & \textbf{32.85} & 11.07 & 23.31 & 23.35 \\
& 31-44 & \textbf{40.85} & 24.75 & \textbf{43.26} & 36.20 \\
& 45-65 & 19.61 & \textbf{34.30} & 17.09 & 23.04 \\
& > 64 & 1.56 & \textbf{24.16} & 15.22 & 13.23 \\
\midrule
\multirow{4}{*}{\textbf{Distance (mi)}} 
& 0-1 & 3.19 & 9.22 & \textbf{86.52} & 32.47 \\
& > 1-3 & \textbf{41.27} & 33.75 & 13.14 & 29.64 \\
& > 3-10 & \textbf{47.99} & 21.89 & 0.34 & 24.12 \\
& >10 & 7.55 & \textbf{35.14} & 0.00 & 13.77 \\
\midrule
\multirow{5}{*}{\parbox{3cm}{\textbf{General Trip purpose}}} 
& Home-based work & \textbf{39.95} & 8.21 & 5.55 & 18.32 \\
& Home-based shopping  & 6.79 & \textbf{30.98} & \textbf{47.65} & 27.55 \\
& \parbox{4cm}{Home-based social/recreational } & 3.60 & \textbf{19.23} & 13.51 & 10.59 \\
& Home-based other & \textbf{33.96} & 23.78 & 13.96 & 26.12 \\
& Not a home-based trip  & 15.70 & 17.80 & 19.33 & 17.43 \\
\midrule
\multirow{4}{*}{\textbf{Race}} 
& White & 34.65 & \textbf{67.59} & 46.95 & 48.47 \\
& Black or African American & \textbf{41.03} & 17.80 & 25.80 & 28.54 \\
& Asian & \textbf{13.20} & 7.28 & 11.24 & 10.52 \\
& Other race & 11.12 & 7.33 & \textbf{16.01} & 12.47 \\
\midrule
\multirow{2}{*}{\textbf{Gender}} 
& Male & 39.99 & 27.53 & \textbf{53.02} & 40.51 \\
& Female & \textbf{60.01} & \textbf{72.47} & 46.98 & 59.49 \\
\midrule
\multirow{2}{*}{\textbf{Employment status}} 
& Worker & \textbf{83.09} & 40.61 & 46.20 & 58.10 \\
& Not worker & 16.91 & \textbf{59.39} & 53.80 & 41.90 \\
\midrule
\multirow{4}{*}{\textbf{Education}} 
& Upto Highschool & 58.32 & \textbf{69.28} & 40.54 & 55.48 \\
& Some College or Trade School & 12.75 & 14.52 & \textbf{27.51} & 17.86 \\
& Bachelor's Degree & \textbf{23.15} & 8.59 & 18.91 & 17.35 \\
& Advanced Degree & 5.79 & 7.62 & \textbf{13.03} & 9.31 \\
\midrule
\multirow{3}{*}{\parbox{3.5cm}{\textbf{Population density (persons per square mile)}}} 
& 0-999 & 12.54 & \textbf{17.93} & 4.69 & 11.59 \\
& 1,000-9,999 & 24.77 & \textbf{73.61} & 52.35 & 48.74 \\
& >10000 & \textbf{62.68} & 8.46 & 42.96 & 39.67 \\
\midrule
\multirow{3}{*}{\textbf{Income (\$)}} 
& 0-24,999 & 40.33 & \textbf{56.36} & 54.95 & 50.50 \\
& 25,000-74,999 & \textbf{36.76} & 22.69 & 24.39 & 27.84 \\
& >74,999 & \textbf{22.90} & 20.96 & 20.66 & 21.66 \\
\bottomrule
\end{tabular}}
\vspace{1ex}
\begin{minipage}{\textwidth}
\small\textit{Note:} Bold values indicate the highest proportion within each variable category across all three classes.
\end{minipage}
\end{table*}

The following provide detailed characterizations of the three identified latent classes based on their modal preferences, trip characteristics, and socio-demographic compositions.

\subsubsection{Class 1: Shared Mobility Errand-Workers}
This segment represents urban-oriented individuals who demonstrate a strong preference for shared mobility (68.85\%) and moderate use of active transportation (22.21\%), with minimal private vehicle dependency (8.94\%). This segment exhibits the highest concentration of employment-related travel, with home-based work trips constituting 39.95\% of all journeys, supplemented by home-based other trips (33.96\%). The employment focus is reinforced by the highest worker participation rate across all classes (83.09\%).

This class is located in high-density urban environments (>10,000 persons/sq mi: 62.68\%), facilitating access to shared mobility infrastructure. The demographic profile is characterized by working age adults, with the highest representation in the 31-44 age range (40.85\%) and some presence of young adults aged 18-30 (32.85\%). Educational attainment shows the largest proportion having completed only high school (58.32\%) yet also displaying moderate rate of bachelor's degree holders (23.15\%) among all classes.

Trip distance patterns reflect intermediate-range mobility needs with medium-distance trips (>3-10 mi: 47.99\%) and short trips (>1-3 mi: 41.27\%) dominating travel behavior. Income distribution spans middle-income brackets (\$25,000-74,999: 36.76\%) more than other classes, though a notable proportion remains in lower income categories (\$0-24,999: 40.33\%). Gender composition favors females (60.01\%), while racial composition shows the highest concentration of Black or African American individuals (41.03\%).

\subsubsection{Class 2: Car-Based Shoppers}
This group exhibits the strongest automobile dependency (81.44\%) and is distinguished by non-work travel purposes, particularly home-based shopping (30.98\%) and social/recreational activities (19.23\%). This class demonstrates the longest average trip distances, with journeys >10 mi representing the highest proportion (35.14\%) across all segments, followed by medium-distance trips (>3-10 mi: 21.89\%).

The demographic profile skews toward older adults, with pre-seniors (45-65: 34.30\%) and seniors (>64: 24.16\%) comprising the majority. This age distribution aligns with the lowest employment rate (40.61\%) and highest proportion of non-workers (59.39\%). Educational level is predominantly at the high school level or below (69.28\%), with limited higher education representation.

Geographically, this segment is concentrated in medium-density suburban areas (1,000-9,999 persons/sq mi: 73.61\%), consistent with car-dependent residential patterns. The class exhibits the highest proportion of white individuals (67.59\%) and maintains a strong female majority (72.47\%). Income characteristics show the highest concentration in the lowest bracket (\$0-24,999: 56.36\%), potentially reflecting fixed incomes among older adults or part-time employment patterns.

\subsubsection{Class 3: Active Travel Shoppers}
This segment demonstrates the most sustainable transportation behavior, with active travel modes (walking, cycling) accounting for 92.37\% of all trips. This segment is characterized by hyperlocal mobility patterns, with 86.52\% of trips occurring within 1 mile of origin, indicating neighborhood-scale activity patterns. Trip purposes are dominated by home-based shopping (47.65\%) and non-home-based trips (19.33\%), suggesting local commercial engagement and chain-trip behavior.

The demographic composition shows a relatively balanced age distribution among working-age cohorts, with adults aged 31-44 (43.26\%) and young adults 18-30 (23.31\%) representing the majority. Unlike other classes, this segment exhibits a male majority (53.02\%) and displays the most diverse racial composition, with representation across white (46.95\%), Black or African American (25.80\%), and Other race (16.01\%) categories.

Spatial distribution favors medium-density (1,000-9,999 persons/sq mi: 52.35\%) and high-density (>10,000 persons/sq mi: 42.96\%) environments that support walkable neighborhood designs. Educational level shows the highest proportion with upto highschool (40.54\%). Employment status is mixed, with a slight majority of non-workers (53.80\%), though this may reflect flexible work arrangements compatible with local activity patterns. Income distribution remains concentrated in lower brackets (\$0-24,999: 54.95\%), potentially reflecting urban residents with lower transportation costs due to reduced vehicle dependency.

\section{Discussion and Implications}
\label{sec:discussion}
 This study reveals that ZVHs employ a range of adaptive strategies to meet their travel needs, shaped by mode availability, trip purpose, demographic characteristics, trip distances, and neighborhood contexts. Through a latent class cluster analysis approach, we identified three distinct behavioral segments—Shared Mobility Errand-Workers, car-based shoppers, and active travel shoppers—each reflecting a unique travel profile in the absence of a household vehicle.
Contrary to the common assumption that ZVHs rely primarily on walking or transit, our findings highlight a more complex reality. Shared Mobility Errand-Workers are predominantly working-age adults who rely heavily on shared mobility options and transit/ridehailing for medium-distance work commutes and essential errands. This segment shows the highest employment rates and. Car-Based Shoppers, though officially car-less, rely extensively on private vehicles likely through informal arrangements, with this segment skewing older, more female, and White. They undertake longer-distance trips, particularly for shopping purposes. Active Travel Shoppers are highly localized, with the vast majority of trips under one mile, relying overwhelmingly on walking and cycling for nearby shopping and demonstrating the highest educational attainment levels.
The probability of belonging to each segment varies dramatically by demographic characteristics. Shared mobility errand-workers are disproportionately Black or African American and represent the most economically constrained yet employment-active segment. Car-based shoppers are White with lower employment rates but greater access to informal vehicle arrangements, suggesting stronger social capital or family networks. Active travel shoppers show markedly higher educational attainment, with more college-educated individuals compared to the car-based shoppers segment.
Environmental context further stratifies these patterns. While active travel shoppers are distributed across urban and suburban density levels, Shared mobility errand-workers are heavily concentrated in the highest-density areas, reflecting both transit availability and housing affordability constraints. Car-based shoppers are most in low and medium-density areas, suggesting that car-less households maintain vehicle access through borrowing than formal ownership.

\subsection{Policy Implications}
\label{subsec:policy_implications}
Our findings emphasize that mobility interventions targeting ZVHs must be differentiated by user segment. A uniform mobility strategy is unlikely to effectively serve such a diverse population with distinct travel patterns, demographic profiles, and spatial distributions.

Active travel shoppers would benefit most from investment in pedestrian and cycling infrastructure across all density levels. These households rely heavily on proximity so policies that encourage mixed-use development and shorter travel distances are essential. Supporting their mobility not only improves access but also promotes environmentally sustainable behavior.

Shared Mobility Errand-Workers require robust shared mobility networks and investments should prioritize high-density urban areas where this segment is concentrated with particular attention to transit coverage, frequency, and affordability.

Car-based shoppers though technically without household vehicles still rely on private vehicle access for loner trips. These individuals might benefit from formalized carsharing programs, neighborhood vehicle cooperatives, or flexible vehicle access schemes. The presence of such services may reduce reliance on informal or costlier vehicle arrangements.

\subsection{Limitations and Future Research}
\label{subsec:limitations}

This study, while offering new insights into the travel behaviors of ZVHs, is not without limitations. Our analysis is based on cross-sectional data which restricts our ability to observe how mobility patterns shift over time. As life circumstances evolve through aging, employment changes, or relocation so too might class membership. Longitudinal research designs are needed to capture these transitions and better understand how individuals adapt over time.

While behavioral patterns are observed, the underlying motivations, preferences, and structural constraints remain unexplored. The private vehicle use among officially car-less car-based shoppers, for example, raises questions about informal access arrangements and choice versus constraint that require qualitative investigation through interviews or focus groups.

\section{Conclusion}
\label{sec:conclusion}

This study employed latent class cluster analysis to identify three distinct travel behavior segments among ZVHs: Shared mobility errand-workers, Car-based shoppers, and active travel shoppers. Each segment demonstrated unique mode preferences, trip patterns, and socio-demographic characteristics revealing the complexity of transportation adaptations in the absence of household vehicles. Key findings include the reliance on shared mobility for work commutes, informal private vehicle use for longer discretionary trips, and active travel for hyperlocal needs, all influenced by factors such as income, age, population density, and trip distance.

The results highlight the importance of tailored transportation policies including investments in shared mobility infrastructure for high-density areas, pedestrian and cycling improvements to support neighborhood-based travel, and programs to formalize vehicle access for car-dependent subgroups. Future research could expand on these insights through longitudinal data or qualitative studies to better understand behavioral motivations. By addressing the diverse needs of ZVHs, planners and policymakers can enhance the efficiency and accessibility of transportation systems for this population.

\section*{Funding}
This research was partly funded by the U.S. Department of Education through the HBCU Master’s Program Grant, by Grant No. P120A210048, the U.S. Department of Transportation’s University Transportation Centers Program grant administered by the Transportation Program at South Carolina
State University (SCSU) and NSF Grants Nos. 2131080, 2242812, 2200457, 2234920, 2305470.

\section*{Acknowledgment}
The authors would like to thank all co-authors and the anonymous reviewers whose comments greatly improved this manuscript.

\section*{Conflict of Interest}
The authors declare that they have no known competing financial interests or personal relationships that could have appeared to influence the work reported in this paper.

\bibliographystyle{ieeetr}
\bibliography{sample}
\end{document}